# The direct Cu NQR Study of the Stripe Phase in the Lanthanum Cuprates

G.B.Teitel'baum[a], B.Büchner[b], H.de Gronckel[c]

[a]Institute for Technical Physics, 420029 Kazan, Russia
[b]II Physikalisches Institut, Universität zu Köln, D-50937 Köln, Germany
[c]IFF Forschungszentrum Jülich, D-52425 Jülich, Germany

Using Cu NQR in Eu-doped $La_{2-x}Sr_xCuO_4$ we find the evidence of the pinned stripe phase at 1.3K for $0.08 \leq x \leq 0.18$. The pinned fraction increases by one order of magnitude near hole doping $x=1/8$. The NQR lineshape reveals three inequivalent Cu positions. A dramatic change of the NQR signal for $x > 0.18$ correlating with the onset of bulk superconductivity corresponds to the depinning of the stripe phase.

## 1. INTRODUCTION

The doping of the antiferromagnetic (AF) insulating phase of a high-$T_c$ superconductor by holes has an explicit topological character. It results in the segregation of charges to the periodical domain walls (stripes) separating the antiphase AF domains [1-3]. The first evidence for such stripe phase has been provided by neutron studies of the low temperature tetragonal (LTT) phase of Nd-doped $La_{2-x}Sr_xCuO_4$ [3]. Recently the stripe correlations were observed in the other cuprates [4-6]. But in spite of the hot interest to the problem suprisingly little is known about the local properties of the stripe structure.

We report the results of the direct study of the stripe phase local structure by means of Cu NQR. The serious difficulties of the present problem are due to the slowing of the charge fluctuations down to MHz frequency range which wipes out a large part of the nuclei from the resonance [7,8]. Fortunately the reappearance of the signal in the slow fluctuations limit at low temperatures enables us to take the advantages of the extreme sensitivity of Cu NQR to the local charge and magnetic field distribution. In addition to the measurements at temperatures of 1.3 K, our program could be realized easier for the LTT structure, which is helpful for pinning of the stripe structure. This structure was induced by doping with non-magnetic Eu rare-earth ions instead of magnetic Nd ones (the ordering of Nd moments causes fast Cu nuclear relaxation hindering the observation of Cu NQR). We expect that in the stripe structure the different Cu sites will be inequivalent with respect to the NQR, providing information on the local properties at given points of the structure.

## 2. EXPERIMENTAL

For our experiments we have chosen fine powders $La_{2-x-y}Eu_ySr_xCuO_4$ with variable Sr content $x$ and fixed Eu content $y=0.17$. The preparation of single-phase samples was described in [9]. It was found [9] that for such Eu content the LTT phase is realized for $x>0.07$. For Sr concentrations $x>0.12$ the ac-susceptibility and microwave absorption measurements reveal the presence of superconductivity with $T_c$ = 6; 9; 14; 19; 18; 16; 13K for resp. $x$= 0.12; 0.13; 0.15; 0.18; 0.20; 0.22; 0.24. The superconducting fraction is small for $x \leq 0.18$ and starting from $x>0.18$ a transition to bulk superconductivity take place.

The NQR measurements are performed with the standard spectrometer in the range 20 - 100 MHz. By lowering the temperature down to 1.3K we succeed to observe the Cu-NQR spectra at all Sr concentrations.

Regarding their NQR properties the samples should be separated into two groups:

The first one corresponds to Sr concentrations $x \leq 0.18$. The superconducting fraction of these samples, if any, was rather small. Each of the spectra, which are very similar for $0.08 \leq x \leq 0.18$, consists of a broad line in the region from 20 MHz up to 80 MHz with an unresolved peak between 30 and 40 MHz (Fig. 1). The spectra for different x differ mainly with the integral intensity, which is peaked near $x=0.12$ (Fig. 2a).

The second group of samples with $x>0.18$ showing bulk superconductivity possesses completely different and much narrower NQR spectra (Inset to Fig.1), which can also be observed at much higher temperatures. The intensity grows up with increasing $x$ from 0.18 (Fig. 2b).

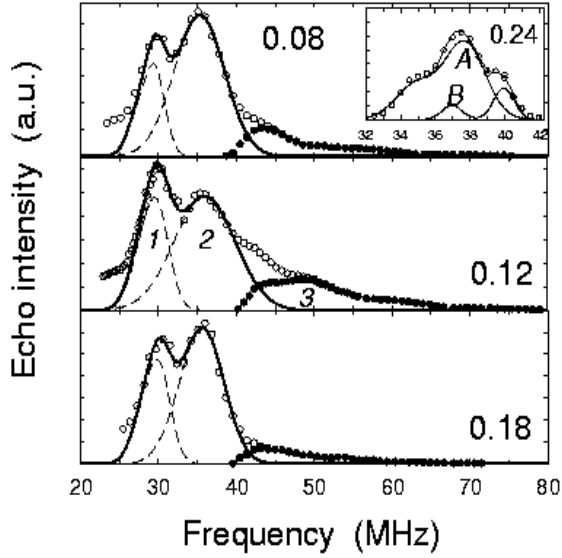

Figure 1. Representative Cu NQR lineshapes at 1.3K of $La_{2-x-y}Eu_yS_xCuO_4$ with $y=0.17$. The value of $x$ is shown for each line. All lineshapes include standard frequency corrections of $v^2$ and are normalized to equal heights. The continuous line is the fit of two-isotope contribution of sites *1* and *2*. Filled circles show the contribution of antiferromagnetic site *3*. Inset: A typical signal for $x>0.18$ decomposed into two contributions (T=4.2K).

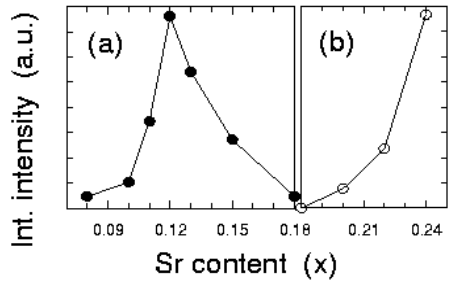

Figure 2. The Cu NQR integrated intensity (normalized to maximal values) of $La_{2-x-y}Eu_yS_xCuO_4$: for $0.08 \leq x \leq 0.18$ at 1.3K (a); for $x \geq 0.18$ at 4.2K (b).

## 3. RESULTS AND DISCUSSION

Beginning the discussion with the $x \leq 0.18$ group, we first consider the above-mentioned complicated peak in the lineshapes. The two Cu-isotopes gaussian fit to these peaks reveals the existence of two independent copper sites *1* and *2* having different NQR frequencies (Fig. 3).

To make the site assignment, we note that the NQR frequency is sensitive to the local hole concentration changing between 0.5 and 0 hole per Cu atom [3]. In a linear approximation we obtain that for the given $x$ the resonant frequency $v_Q$ is connected with the local hole density $n(\mathbf{r})$ via the relation $v_Q(x,n) = v_Q^0 - \alpha x + \beta n$ with the empirical constants $\alpha$ and $\beta$. The first term here is the NQR frequency for the compound with zero Sr content, the second one is due to the negative shift caused by the contraction of Cu-O bond length upon substitution of La with Sr, the third corresponds to the positive shift due to the local increase of the effective fractional charge on Cu. This expression agrees both with the calculations in the ionic [10] as well as in the cluster [11] models (in the uniform case $n=x$).

It follows from our results (Fig.3) that the resonance frequencies $^{63}v_Q^{(1)}(x)$ for line *1* are shifted to lower values from the reference value $^{63}v_Q(0,0)=^{63}v_Q^0$ (we use here $^{63}v_Q^0 = 31.9$MHz estimated for $La_2CuO_4$ [12]). This indicates that the positive contribution to $^{63}v_Q(x,n)$ is small and that the effective fractional charge on sites *1* is near zero. It means that they are located in the regions free of doped holes. In contrast line *2* is due to the sites which in addition to the negative shift exhibit a positive one. It means that these sites belong to the regions with an increased average charge (hole density) on the Cu ions. The high frequency part of the spectrum can be analyzed by subtraction of the *1* and *2* contributions from the entire signal. The resulting spectra are shown in Fig.1. The frequencies corresponding to their maxima are plotted in Fig.3. We assume that this line corresponds to the broadened $+1/2 \leftrightarrow -1/2$ transitions of nuclei located in sites *3* experiencing an internal magnetic field (note the broad high-frequency shoulder). The satellites are unresolved due to inhomogeneities of the internal magnetic field and of the NQR frequencies. If the orientation of the internal field with respect to the electric field gradient is identical to that observed for $La_2CuO_4$ [12] the frequency of this transition enables us to estimate the quadrupole shift and to determine the Larmor frequency for this Cu site to be 45.2 MHz for $x=0.12$. It corresponds to an internal field of 40.1 kOe. Using the hyperfine constant $|A_Q|=139$ kOe/$\mu_B$ [13] we estimate the effective magnetic moment of Cu at site *3* to be equal to 0.29 $\mu_B$, coinciding with the value obtained from neutron and muon experiments [4,14].

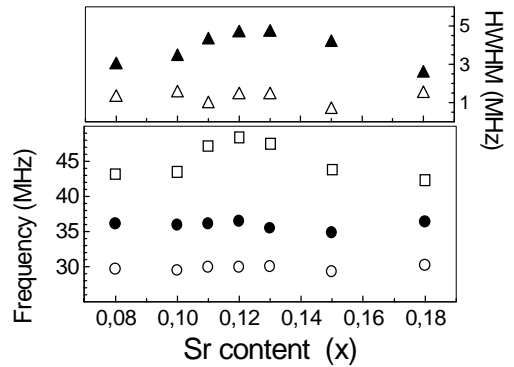

Figure 3. The parameters of the different contributions to the Cu NQR signal for $0.08 \leq x \leq 0.18$ in $La_{2-x-y}Eu_yS_xCuO_4$ (T=1.3K): $^{63}Cu$ NQR frequencies of sites *1* - open circles and *2* - filled circles; the corresponding half widths at half maximum (HWHM) - open and filled triangles resp.; the frequencies corresponding to the maxima of the magnetic contribution *3* are shown by squares.

Since quantitatively similar spectra were observed for each compound of the first group we believe that they contain the same elementary "bricks" of the phase under study. Discussing the relative weight of the different contributions, we note that the echo decay can be described in terms of stretched exponents $exp[-(2t/T_2)^a]$ with $a \cong 0.5$ and different $T_2$ for each site. For $x=0.12$: $T_2^{(1)} \cong 11\mu sec$; $T_2^{(2)} \cong 8.8\mu sec$; $T_2^{(3)} \cong 5.5\mu sec$. Extrapolating the corresponding signal intensities to $t=0$ we find the contributions of sites *1*, *2*, *3* to be given by the ratio (1:6:13).

As for the origin of the sites *1* it is possible to conclude that on one hand they do not belong to the AF domains, and on the other hand they are outside of stripes since their effective charge is nearly zero. We assume that they correspond to defects terminating the stripes. From their relative number we estimate the average length of the stripe to equal at least 6 lattice constants. Note, that since the Eu doping produces the contraction of Cu-O bonds [9] and causes the negative shift of the Cu NQR frequencies, there is also a certain contribution to line *1* of Cu sites underneath a Eu atom creating the enhanced tail at the low frequency side of the observed spectra.

The NQR frequencies for the site *2* (See Fig.4) are almost the same for any $x$ thus indicating that for all Sr concentrations the stripes are equally charged. The effective charge in a stripe is near 0.18-0.19. This is larger than the average hole concentration ($x$) but less than 0.5 expected for the ideal stripe picture [3]. It means that the charge is distributed over the domain wall of a finite thickness. Together with the above-mentioned intensity ratio this indicates that the real stripe picture differs from the ideal one. The changes in intensity of the NQR spectra are due to variation of the number of "bricks" for the compounds with different $x$, which depends on the pinning strength. Our results indicate, that the stripe phase is pinned at least for the time scales shorter than $10^{-6}$ sec.

The pinning for $0.08 \leq x \leq 18$ is due to the buckling of the $CuO_2$ plane. It is connected with the $CuO_6$ octaedra tilts around the [100] and [010] axis by the angle $\Phi$, which for given Eu substitution is governed by the Sr content. It follows from Fig. 2a, that the quantity of the pinned phase, which is proportional to the NQR signal intensity, is peaked at $x=0.12$ when according to our estimations the entire sample is in the pinned stripe phase. This indicates additional strong pinning due to the commensurability effect. Such pinning is not unique for the LTT phase (as buckling is). It is a manifestation of the plane character of the inhomogeneities of the charge and spin distributions. Together with the existence of three different Cu sites (note that site *1* corresponds to a defect) this gives an independent justification of the stripe picture. Our data are compatible with two models: a) the structure [3] with the charge domain wall along one Cu-O chain separated by the bare three-leg ladders. The deviation of the estimated effective charge from 0.5 may indicate that in reality the thickness of the charged stripe is larger than one lattice spacing; b) the structure [15] with the charged walls consisting from two neighboring Cu-O chains separated by two-leg spin ladders. The average hole density in this model is 0.25. This is closer to the observed 0.19, but there are some problems with the relative intensity of the sites *2* and *3*.

Upon increasing $x$ over $x=0.18$ the tilt angle is decreasing below the critical value $\Phi_c \cong 3.6^o$ [16] and depinning of the stripe phase takes place. Such behaviour occurs for the compounds with Sr concentration $x>0,18$. The corresponding NQR spectrum transforms to the narrow signal at higher frequencies, which for $x=0.24$ is shown in the inset to Fig. 1. The intensity of this line (proportional to the quantity of the unpinned stripe phase) is shown in Fig. 2b.

The analysis of this relatively narrow signal reveals only two different sites with $^{63}$Cu NQR frequencies of 37.60 MHz and 39.82 MHz. Within 1% accuracy these frequencies coincide with those known at the same $x$ for the A and B sites in the LTO superconducting phase [17] confirming that the LTT structure differs only in the directions of CuO$_6$ octaedra tilts. The satellite B is due to Cu having a localized hole in the nearest surrounding since, according to [11, 18], its NQR frequency has the additional positive shift $\delta\nu_Q \cong 2.5$MHz. The observed transformation of the NQR spectra in comparison with those for $x \leq 0.18$ is due to the fast transverse motion of stripes in the depinned phase. As a result the internal magnetic field on Cu nuclei is averaged out, and the effective fractional charge is homogeneously distributed over all Cu nuclei giving the usual NQR frequencies. Such depinning leads to the drastic changes in the magnetic properties. The echo signals decay for samples with $x>0.18$ becomes purely exponential $T_2^{(2)} \cong 35.4\mu$sec for sample with $x=0.24$.

Regarding the superconducting properties we note that the depinning point separates two different types of superconductivity. For $x \leq 0.18$ we are dealing with a weak Meissner effect, an increased London penetration length and with $T_c$ increasing with $x$ growing up to 0.18. Combining these facts with the absence of a narrow signal typical for the bulk superconducting phase, indicating that the impure LTO phase is absent, and with the suppression of the relaxation via magnetic moments of doped holes, one has arguments in favor of possible one-dimensional superconductivity along the charged rivers of stripes - the issue which is widely discussed [19]. For $x>0.18$ we have bulk superconductivity with conventional London length, typical NQR signal and decreasing $T_c(x)$. Such crossover may be caused by the transverse motion of the stripes carrying superconducting currents, which gives rise to the conventional superconductivity in CuO$_2$ planes. Although possibly a simple coincidence, it happens when the doping $x$ is equal to the effective charge ($n$) in a stripe.

## 4. CONCLUSIONS

The Cu NQR of the Eu doped La$_{2-x}$Sr$_x$CuO$_4$ was studied. We demonstrated that at 1.3K the ground state for moderate Sr content corresponds to the pinned stripe-phase and that the pinning is enhanced at the commensurability. Three nonequivalent copper positions in the CuO$_2$ planes were found. One of them with a magnetic moment of 0.29$\mu_B$ is related to the AF correlated domains. From the behaviour of the NQR frequencies it follows that the effective charge of the domain walls separating these domains is almost independent on the Sr content $x$. The onset of the bulk superconductivity at larger $x$ correlates with the dramatic transformation of the NQR spectra, indicating the depinning of the stripe phase.

## 4. ACKNOWLEDGEMENTS

The authors are grateful to H.Brom, N.Garifjanov A.Egorov, M.Vojta and S. Sachdev for valuable discussions. The work of G.T. was supported in part by the State HTSC Program of the Russian Ministry of Sciences (Grant No. 98001) and by the Russian Foundation for Basic Research (Grant No. 98-02-16582).